\documentclass[a4paper,10pt]{article}
\usepackage[utf8]{inputenc}
\usepackage{authblk}
\usepackage[style=authoryear,giveninits=true,url=false,doi=false]{biblatex}
\DeclareNameAlias{author}{family-given}
\usepackage{csquotes}
\usepackage{parkers}
\usepackage[title]{appendix}

\DeclareNameFormat{labelname:poss}{%
  \ifcase\value{uniquename}%
    \usebibmacro{name:family}
      {\namepartfamily}
      {\namepartgiven}
      {\namepartprefix}
      {\namepartsuffix}%
  \or
    \ifuseprefix
      {\usebibmacro{name:given-family}
        {\namepartfamily}
        {\namepartgiveni}
        {\namepartprefix}
        {\namepartsuffixi}}
      {\usebibmacro{name:given-family}
        {\namepartfamily}
        {\namepartgiveni}
        {\namepartprefixi}
        {\namepartsuffixi}}%
  \or
    \usebibmacro{name:given-family}
      {\namepartfamily}
      {\namepartgiven}
      {\namepartprefix}
      {\namepartsuffix}%
  \fi
  \ifnumequal{\value{listcount}}{\value{liststop}}
    {'s}
    {}%
  \usebibmacro{name:andothers}}

\newrobustcmd*{\posscitealias}{%
  \AtNextCite{%
    \DeclareNameAlias{labelname}{labelname:poss}}}

\newrobustcmd*{\posscite}{%
  \posscitealias
  \textcite}

\newrobustcmd*{\Posscite}{\bibsentence\posscite}

\newrobustcmd*{\posscites}{%
  \posscitealias
  \textcites}

\title{Polytropic Gas Effects in Parker's Solar Wind Model and Coronal Hole Flows}
\author[1,2]{B. K. Shivamoggi}
\author[2]{L. Pohl}
\affil[1]{California Institute of Technology, Pasadena, CA 91125}
\affil[2]{University of Central Florida, Orlando, FL 32816}

\begin{document}

\maketitle

\begin{abstract}
A detailed and systematic investigation of polytropic gas effects in Parker's solar wind model and coronal-hole flows is given. We present a viable equation governing the acceleration of solar wind of a polytropic gas and give its \emph{exact analytical} and numerical solutions and deduce its asymptotic analytic properties (i) near the sun, (ii) far away from the sun, (iii) near the Parker sonic critical point (where the wind speed is equal to the speed of sound in the wind). We proceed to give a detailed and systematic investigation of coronal-hole polytropic gas outflows which contribute to bulk of the solar wind. We will model coronal-hole outflow by considering a single radial stream tube and invoke phenomenological considerations to represent its rapidly-diverging flow geometry. We give analytical and numerical solutions for this outflow and deduce its asymptotic analytic properties in the three flow regimes above. We find that, in general, the polytropic effects cause the Parker sonic critical point to move closer to the sun than that for the case with isothermal gas. Furthermore, the flow acceleration is found to exhibit (even for an infinitesimal deviation from isothermality of the gas) a \emph{power-law} behavior rather than an exponential-law behavior near the sun or a logarithmic-law behavior far away from the sun, thus implying a certain robustness of the power-law behavior. The Parker sonic critical point is shown to continue to be of $X$-type, hence facilitating a smooth transition from subsonic to supersonic wind flow through the transonic regime. Our analytical and numerical solutions for coronal-hole outflows show that the super-radiality of the stream tube causes the Parker sonic critical point to move further down in the corona, and the gas to become more diabatic (the polytropic exponent $\gamma$ drops further below $5/3$), and the flow acceleration to be enhanced further.
\end{abstract}
\newpage
\section{Introduction}
Solar wind is a continuous plasma outflow from the sun, the bulk of which emerges from the coronal holes \parencite{Sakao2007}. Solar wind carries off a huge amount of angular momentum from the sun while causing a negligible mass loss from the sun. Weak to moderate solar winds are produced by an extended active heating of the corona in conjunction with high thermal conduction. \textcite{Parker1958} pointed out that high-speed solar winds need some additional acceleration mechanism operational beyond the coronal base and gave an ingenious model to continually convert the coronal thermal energy into the kinetic energy of the wind to produce a smooth acceleration of the latter through transonic speeds (i.e., wind speeds near the speed of sound in the wind). Recent \emph{in situ} measurements \parencite{Rivera2024} of a well-identified plasma patch from two solar space probes (Parker Solar Probe and Solar Orbiter) indicated that heating and acceleration experienced by this patch seemed to be traceable to the magnetic activity associated with this pathc. The solar wind and its properties have been recorded by the \emph{in-situ} observations \parencite{Meyer-Vernet2007}. Incoming data from the Parker Solar Probe has been providing considerable useful information about the solar corona conditions (\textcite{Fisk2020} and the references thereof), some of which (like the coupling of the solar wind with the solar rotation \parencite{Kasper2019}) can be the cause of faster solar winds \parencite{Shivamoggi2020} via the mechanism of centrifugal driving.

A major assumption in Parker's original solar wind model \parencite{Parker1958} was that the gas flow occurs under isothermal conditions, i.e., (in standard notation),
\begin{align*}
 p &= a^2\, \rho\,,
\end{align*}
where the speed of sound $a$ is taken to be constant. On the other hand, the solar wind is found \parencite{Boldyrev2020} not to cool down as fast as that caused even by an adiabatic expansion, indicating the presence of significant heating in the corona impairing adiabaticity. A polytropic gas model \parencite{Parker1971, Holzer1979, Shivamoggi2021}, described by
\begin{align*}
 p &= \mathcal{C}\, \rho^\gamma\,,
\end{align*}
where $\gamma$ is the polytropic gas exponent, $1<\gamma<5/3$, and $\mathcal{C}$ is an arbitrary constant, is suitable for this situation. The formulation given by \textcite{Holzer1979} is not in a convenient form and does not facilitate a systematic exploration of polytropic gas effects as well as easy comparison with the isothermal gas resuts treated by \textcite{Parker1958}. On the other hand, the formulation of \textcite{Shivamoggi2021} did not include the effects of solar gravity in estimating the stagnation flow properties in the solar wind and is hence quantitatively not very accurate near the coronal base. The purpose of this paper is to remedy these issues, and to provide a detailed and systematic investigation of polytropic gas effects on the dynamics of the solar wind. Toward this objective we present a viable equation governing the acceleration of solar wind of a polytropic gas and give its \emph{exact analytical} and numerical solutions and deduce its asymptotic analytic properties,
\begin{itemize}
 \item near the sun,
 \item far away from the sun,
 \item near the Parker sonic critical point (where the wind speed is equal to the speed of sound in the wind).
\end{itemize}
Following \textcite{Bondi1952}, we formulate the solar wind model for the polytropic gas by introducing the Mach number $M$ of the flow $M \equiv v_r/a$, which turns out to be the optimal variable for describing the polytropic gas flow.

The coronal-hole outflows, which contribute to the bulk of the solar wind, are known to be permeated radially by divergent open unipolar magnetic field lines threading them \parencite{Altschuler1972,Krieger1973,Nolte1976}. Furthermore, coronal-hole outflows are found to be super-radial involving rapidly divergent geometries in the sense that the cross-sectional area of a given stream tube in this inner-corona region increases outward from the sun faster than $r^2$ (that corresponding to a spherical geometry). Recent Parker Solar Probe observations \parencite{Bale2023} indicate the \emph{fast} solar wind emerges from the coronal holes via the process of magnetic reconnection between the open and closed magnetic field lines (called the \emph{interchange connection}). On the other hand, recent \emph{in situ} observations from Solar Orbiter \parencite{Yardley2024} indicated latter process to be the cause of the \emph{slow} solar wind as well. \textcite{Kopp1976,Wang1990} proposed to model the coronal-hole outflows by suitable super-radial geometries, and the numerical calculations of \textcite{Kopp1976} showed enhancement of the flow acceleration and hence a transition to supersonic flow closer to the sun.

In order to circumvent the mathematical difficulties posed by an explicit consideration of non-radial flows, we will model a coronal-hole outflow by a magnetic-field aligned infinitesimal radial stream tube, and consider only a single stream tube because it would disrupt radial geometry for the neighboring stream tubes. We may then, on phenomenological grounds, represent the rapidly-diverging geometry of the stream tube by taking the cross-sectional area $A(r)$ to increase outward from the sun faster than $r^2$ (that corresponding to a spherical geometry), as described by $A(r) \sim r^\beta$, $\beta>2$. We will use this phenomenological model to provide a detailed and viable theoretical investigation of coronal-hole outflows in a polytropic gas. We will present an equation governing such flows and give its \emph{exact analytical} and numerical solutions, and deduce its asymptotic analytic properties in the three regimes listed previously.

\section{Solar Wind Model with a Polytropic Gas}
\posscite{Parker1958} solar wind model assumes a steady and spherically symmetric radial flow so the flow variables depend only on the radial distance $r$ from the sun. Furthermore, the flow variables and their derivatives are assumed to vary continuously so there are no shocks anywhere in the region under consideration.

The equations expressing the conservation of mass and momentum balance are (in usual notations):
\begin{align}
 \frac{2}{r} + \frac{1}{\rho}\ddd{\rho} + \frac{1}{v_r}\ddd{v_r} &= 0\,,\label{eqn:mass_cons}\\
 \rho\,v_r \ddd{v_r} &= -\ddd{p} - \rho\frac{\gm}{r^2}\,,\label{eqn:mome_cons}
\end{align}
where $G$ is the gravitational constant and $M_\odot$ is the mass of the sun.

We assume the polytropic gas relation,
\begin{align}
 p &= \mathcal{C}\,\rho^\gamma\,,\label{eqn:polytrop}
\end{align}
where $\mathcal{C}$ is an arbitrary constant, and $\gamma$ is the polytropic exponent, $1<\gamma<5/3$, which characterizes the extent to which conditions within the solar coronal gas depart form adiabatic conditions ($\gamma = 5/3$) due to coronal heating effects via thermal conduction and wave dissipation \parencite{Parker1960}.\footnote{A more detailed investigation is to make use of a complete energy equation with thermal conduction and heating effects in the solar corona.}

Using equations (\ref{eqn:mass_cons})--(\ref{eqn:polytrop}), we obtain,
\begin{align}
 \left( \frac{v_r^2}{a^2} - 1\right) \frac{2 v_r}{a^2} \ddd{v_r} &= \frac{4}{r^2}\frac{v_r^2}{a^2}\,(r - r_*)\,,\label{eqn:mom2}
\end{align}
$a$ is the speed of sound in the gas, and $r_*$  locates the Parker sonic critical point in the corona where the flow speed equals the local sound speed,
\begin{subequations}
 \begin{align}
 a^2 &\equiv \ddd[\rho]{p}\,,\label{eqn:sound}\\
 r_* &\equiv \frac{\gm}{2 a^2}\label{eqn:rs}\,.
 \end{align}
\end{subequations}

We now follow \posscite{Bondi1952} suggestion, in the context of the related accretion disk model, that the polytropic case is best formulated by introducing the Mach number $M$ of the flow, $M\equiv v_r/a$.

Introducing the total energy $E$, we have,
\begin{align}
 \frac{v_r^2}{2} + \frac{a^2}{\gmj} - \frac{\gm}{r} & = E \equiv \frac{\atd}{\gmj}\label{eqn:eng}\,,
\end{align}
from which, we obtain:
\begin{align}
 \frac{\atd}{a^2} &= \frac{1 + \gmjp M^2}{1 + 4\gmjp \frac{\tir}{r}}\,,\label{eqn:eng2}
\end{align}
where $\tir \equiv \frac{\gm}{2\atd}$.

We have,
\begin{align}
 \ddd{M^2} &= \frac{2 v_r}{a^2}\ddd{v_r} - \frac{2 v_r^2}{a^3}\ddd{a}\,.\label{eqn:mach2}
\end{align}

Using (\ref{eqn:eng2}), (\ref{eqn:mach2}) leads to,
\begin{align}
 \frac{2v_r}{a^2} \ddd{v_r} &= \left[ \frac{1}{1 + \gmjp M^2} \right] \ddd{M^2} - M^2 \left[ \frac{4 \gmjp \frac{\tir}{r^2}}{1 + 4 \gmjp \frac{\tir}{r}}\right]\,.\label{eqn:mom3}
\end{align}
Using (\ref{eqn:mom3}), (\ref{eqn:mom2}) leads to,
\begin{align}
    \begin{split}
    &\frac{M^2-1}{M^2\left( 1 + \gmjp M^2 \right)}\ddd{M^2} - (M^2 - 1)\left[ \frac{4\gmjp\frac{\tir}{r^2}}{1 + 4 \gmjp \frac{\tir}{r}}\right] \\
    &= \frac{4}{r^2} \left( r - \frac{1+\gmjp M^2}{1+ 4\gmjp \frac{\tir}{r}}\;\tir\right)\,,\label{eqn:mom4}
    \end{split}
\end{align}
or on simplification, the equation governing the polytropic gas flow becomes,
\begin{align}
 \frac{M^2-1}{M^2\left( 1 + \gmjp M^2 \right)}\ddd{M^2} &= \frac{4}{r^2} \left( r - \gpjp\frac{1}{1+ 4\gmjp \frac{\tir}{r}}\;\tir\right)\,.\label{eqn:mom5}
\end{align}

Equation (\ref{eqn:mom5}) is the generalization of \textcite{Parker1958} isothermal solar wind flow equation for the polytropic gas, and reduces to the former in the isothermal limit $\gamma = 1$.

\begin{figure}[ht]
 \centering
 \includegraphics[width=\textwidth]{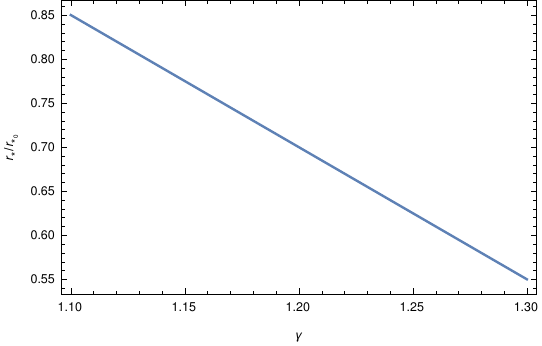}
 \caption{Polytropic effects, as per (\ref{eqn:cp}), on the Parker sonic critical point.}
 \label{fig:r_star}
\end{figure}

Observe from equation (\ref{eqn:mom4}) or (\ref{eqn:mom5}) that the Parker sonic critical point for a polytropic gas is given by,
\begin{align*}
 M^2 = 1\,\text{:}\quad r_* = \gpjp\frac{1}{1+4\gmjp\frac{\tir}{r_*}}\tir =0\,,
\end{align*}
or
\begin{subequations}\label{eqn:cp}
\begin{align}
  M^2 = 1\,\text{:}\quad r_*= \pmtgp\;\tir\,.\label{eqn:cp1}
\end{align}
We may rewrite (\ref{eqn:cp1}) as
\begin{align}
  M^2 = 1\,\text{:}\quad r_*= \left[ 1 - \frac32\left(\gmj\right)\right]\tir < \tir\,,\label{eqn:cp2}
\end{align}
\end{subequations}
which implies that the polytropic effects ($\gamma > 1$) cause the Parker sonic critical point to move closer to the sun (see Figure~\ref{fig:r_star}) than that for the case with isothermal gas ($\gamma = 1$: $r_* = \tir$), in agreement with our previous results \parencite{Shivamoggi2021} for the more restricted polytropic gas case.

\section{Exact Solution}
Upon imposing the boundary condition (\ref{eqn:cp}), equation (\ref{eqn:mom5}) yields the \emph{exact} solution,
\begin{align}
 \begin{split}
 &-\log{M^2} + \gpmj \log{\left( \frac{1+\gmjp M^2}{\gpjp} \right)} \\
 &= \frac{3\gamma - 5}{\gmj}\log{\left(\frac{2r}{(\pmtg)\tir}\right)} + \gpmj\log{\left( \frac{r + 4\gmjp \tir}{\gpjp \tir}\right)}\,.\label{eqn:solex}
 \end{split}.
\end{align}

\begin{figure}[htb]
 \centering
 \includegraphics[width=\textwidth]{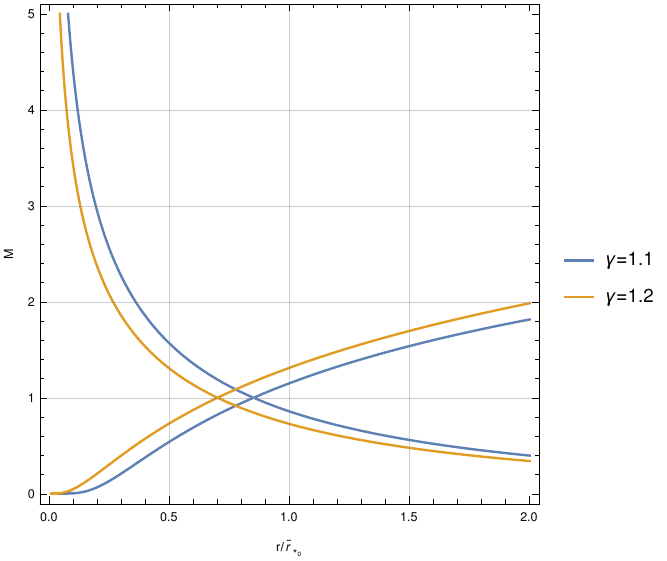}
 \caption{Exact solution (\ref{eqn:solex}) for the polytropic solar wind for two values of polytropic exponent $\gamma$.}
 \label{fig:polytropic_glob_stag}
\end{figure}

In the isothermal gas limit, ($\gamma=1$), (\ref{eqn:solex}) reduces to the Parker solar wind solution,
\begin{align}
 -\log{M^2} + M^2 &= 4\log{\frac{r}{\tir}} + 4\frac{\tir}{r} - 3\,.\label{eqn:solpark}
\end{align}
The exact solution (\ref{eqn:solex}) is plotted in Figure~\ref{fig:polytropic_glob_stag}, for different values of the polytropic exponent $\gamma$. Observe the enhanced acceleration of the solar wind caused by the polytropic gas effects, which is plausible because of enhanced flow expansion in the latter case.

\subsection{Near-sun Regime}
For $r/\tir\ll1$, (\ref{eqn:solex}) reduces to:
\begin{subequations}\label{eqn:nearsun}
\begin{align}
 -\log{M^2} \approx \frac{3\gamma - 5}{\gmj} \log{r}\,,\label{eqn:nearsun1}
\end{align}
from which,
\begin{align}
 M^2 \sim r^{\frac{\pmtg}{\gmj}}\,.\label{eqn:nearsun2}
\end{align}
\end{subequations}
(\ref{eqn:nearsun2}) is valid, provided $1<\gamma<5/3$, which implies the coronal gas needs to be heated to achieve $\gamma<5/3$.

On the other hand, in the isothermal gas limit ($\gamma = 1$), (\ref{eqn:nearsun2}) is not valid, and (\ref{eqn:solpark}) yields:
\begin{align}
 M^2 \sim e^{-4\frac{\tir}{r}}\,.
\end{align}

Thus, any small deviation from the gas isothermality ($\gamma\neq1$) leads to a power-law (instead of an exponential law) enhancement of flow acceleration near the sun, so the power-law behavior given by (\ref{eqn:nearsun2}) needs to be taken as a more robust one. On the other hand, this signifies that the isothermal limit, $\gamma\to1$, is a \emph{singular} limit of the polytropic gas, $\gamma>1$, so the polytropic gas results are not obtainable via a Taylor series expansion about the isothermal case.

\subsection{Far-sun Regime}
For $r/\tir\gg1$, (\ref{eqn:solex}) reduces to:
\begin{subequations}\label{eqn:farsun}
\begin{align}
 \left( -1 + \gpmj\right)\log{M^2} &\approx \left( \frac{3\gamma - 5}{\gmj} + \gpmj\right) \log{r}\,,\label{eqn:farsun1}
\end{align}
from which,
\begin{align}
 M^2 &\sim r^{2(\gmj)}\,.\label{eqn:farsun2}
\end{align}
\end{subequations}
On the other hand, in the isothermal gas limit ($\gamma=1$), (\ref{eqn:solpark}) yields,
\begin{align}
 M^2 &\sim \log{r}\,.
\end{align}
Thus, any small deviation from the gas isothermality ($\gamma\neq1$) leads to a power-law (instead of a logarithmic law) enhancement of flow acceleration far from the sun, so the power-law behavior given by (\ref{eqn:farsun2}) needs to be taken again as a more robust one.

\subsection{Transonic Regime}
Near the Parker sonic critical point, $r = r_* = \pmtgp\,\tir$, we write,
\begin{align}
 r &\equiv \pmtgp\tir + x\,,\quad M^2 \equiv 1 + y\,,
\end{align}
(\ref{eqn:solex}) then reduces to,
\begin{align*}
    \begin{split}
    &-\log{(1+y)} + \gpmj \log{\left( 1+\gmpj\,y\right)} \\
    &= -\frac{\pmtg}{\gmj}\log{\left( 1 + \frac{2}{\pmtg}\frac{x}{\tir}\right)} + \gpmj \log{\left( 1 + \frac{2}{\gpj}\frac{x}{\tir}\right)}\,,
    \end{split}
\end{align*}
and Taylor expanding in powers of $y$ and $x/\tir$, we obtain,
\begin{align*}
 \frac{1}{\gpj}\; y^2 &\approx \frac{1}{\gmj} \left( \frac{2}{\pmtg} - \frac{2}{\gpj}\right)\left(\frac{x}{\tir}\right)^2\,,
\end{align*}
from which,
\begin{subequations}\label{eqn:tran}
\begin{align}
 y &\approx \sqrt{\frac{8}{\pmtg}}\, \frac{x}{\tir}\,,\label{eqn:tran1}
\end{align}
or
\begin{align}
 M^2 - 1 &\approx \sqrt{\frac{8}{\pmtg}}\; \left( \frac{r}{\tir} - \pmtgp\right)\,.\label{eqn:tran2}
\end{align}

(\ref{eqn:tran1}) and (\ref{eqn:tran2}) describe the asymptotes near the Parker sonic critical point ($M^2 = 1$, $r = \pmtgp\,\tir$) to the hyperbolas given by (\ref{eqn:solex}). This result implies that the Parker sonic critical point is of $X$-type, which facilitates a smooth transition from subsonic to supersonic wind flow through the transonic regime ($M\approx1$). On rewriting (\ref{eqn:tran1}) and (\ref{eqn:tran2}) as:
\begin{align}
 y &\approx \sqrt{\frac{8}{2-3(\gmj)}}\, \frac{x}{\tir}\,,\label{eqn:tran3}
\end{align}
or
\begin{align}
 M^2 - 1 &\approx \sqrt{\frac{8}{2-3(\gmj)}}\; \left( \frac{r}{\tir} - \frac{2-3(\gmj)}{2}\right)\,,\label{eqn:tran4}
\end{align}
\end{subequations}
we see the enhancement in the flow acceleration past the Parker  sonic critical point caused by the polytropic gas effects ($\gamma > 1$).

It may be noted that the above asymptotic results (\ref{eqn:nearsun2}), (\ref{eqn:farsun2}), (\ref{eqn:tran2}), for
$r/\tir\ll1$, $r/\tir\gg1$, and $r\approx \pmtgp\,\tir$ may be deduced directly from equation (\ref{eqn:mom4}) governing the polytropic gas flow (see Appendix \ref{sec:app1}).

\section{Coronal Hole Flows}
The super-radial coronal-hole outflows involve rapidly diverging geometries and pose difficulties by considering them explicitly. In order to circumvent these difficulties, we will model the coronal-hole outflow by a magnetic-field aligned infinitesimal radial stream tube, and consider only a single stream tube because it would disrupt radial geometry for the neighboring stream tubes. Furthermore, the cross-sectional area $A(r)$ of the stream tube appears in the equation governing the flow via only the ratio $(1/A(r)) (\dd A/\dd r)$. So, one may, on phenomenological grounds, represent the rapidly-diverging flow geometry of the stream tube by taking the cross-sectional area $A(r)$ to increase outward from the sun faster than $r^2$ (corresponding to a spherical geometry), as described by,
\begin{align}
 A(r) &\sim r^\beta\,,\quad \beta>2\,.\label{eqn:areabeta}
\end{align}
\subsection{Isothermal Gas}
Consider flow of isothermal gas in a stream tube of cross-sectional area $A(r)$. Equation (\ref{eqn:mass_cons}) then becomes:
\begin{align}
 \frac{1}{\rho}\ddd{\rho} + \frac{1}{v_r}\ddd{v_r} + \frac{1}{A}\ddd{A} &= 0\,.\label{eqn:mass_consbeta}
\end{align}
Equations (\ref{eqn:mome_cons}) and (\ref{eqn:mass_consbeta}) lead to
\begin{align}
 \frac{1}{v_r}\left( v_r^2 - a^2 \right) \ddd{v_r} &= \frac{2a^2}{r^2}\left( \frac{r^2}{2A}\ddd{A} - r_*\right)\,.\label{eqn:mom_beta}
\end{align}
Using (\ref{eqn:areabeta}), equation (\ref{eqn:mom_beta}) becomes,
\begin{align}
 \frac{1}{v_r}\left( v_r^2 - a^2 \right) \ddd{v_r} &= \beta\frac{a^2}{r^2} \left( r - \frac{2}{\beta}r_* \right)\,.\label{eqn:mom_beta2}
\end{align}

Observe from equation (\ref{eqn:mom_beta}) that the sonic critical point for a coronal-hole stream tube is given by
\begin{align}
 r &= \frac{2}{\beta}r_*<r_*\;\text{: }\; v_r = a\,,\beta>2\label{eqn:mom_beta_crit}
\end{align}
which implies that the Parker sonic critical point occurs closer to the sun for a coronal-hole outflow ($\beta>2$).

Upon imposing the boundary condition (\ref{eqn:mom_beta_crit}), equation (\ref{eqn:mom_beta2}) yields the exact solution:
\begin{align}
 \frac{v_r^2}{a^2} - \log{\left( \frac{v_r}{a} \right)^2} &= 2\beta \log{\left( \frac{\beta r}{2r_*}\right)} + 4\,\frac{r_*}{r} - (2\beta - 1)\,.\label{eqn:exsolisobeta}
\end{align}

In the radial-flow limit, $\beta=2$, (\ref{eqn:exsolisobeta}) reduces to the Parker solar wind solution (\ref{eqn:solpark}).

\subsection{Polytropic Gas}
Consider next flow of a polytropic gas in a stream tube of cross-sectional area $A(r)$. Using equations (\ref{eqn:areabeta}) and (\ref{eqn:mass_consbeta}), equation (\ref{eqn:mom4}) becomes,
\begin{align}
    \begin{split}
    &\frac{M^2 - 1}{M^2\left( 1 + \gmjp M^2 \right)}\ddd{M^2} - (M^2 - 1)\left( \frac{4\,\gmjp\frac{\tir}{r^2}}{1+4\,\gmjp\frac{\tir}{r}}\right)\\
    &= \frac{4}{r^2}\left[ \frac{\beta}{2}\,r - \left( \frac{1+\gmjp M^2}{1+ 4\,\gmjp\frac{\tir}{r}}\right)\,\tir\right]\,,\label{eqn:mom4beta}
    \end{split}
\end{align}
or on simplification, the equation governing the super-radial polytropic gas flow becomes,
\begin{align}
 \frac{M^2-1}{M^2\left( 1 + \gmjp M^2 \right)}\ddd{M^2} &= \frac{4}{r^2} \left( \frac{\beta}{2}\,r - \gpjp\frac{1}{1+ 4\gmjp \frac{\tir}{r}}\;\tir\right)\,.\label{eqn:mom5beta}
\end{align}

\begin{figure}[ht]
 \centering
 \includegraphics[width=\textwidth]{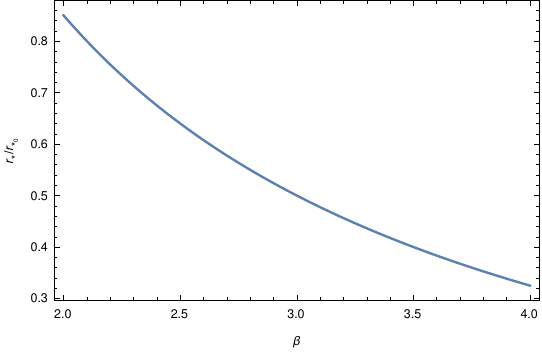}
 \caption{Super-radial flow effects, as per (\ref{eqn:cpbeta}), on the Parker sonic critical point ($\gamma = 1.1$).}
 \label{fig:r_star_beta}
\end{figure}

Observe from equation (\ref{eqn:mom5beta}) that the Parker sonic critical point for a coronal-hole stream tube is given by,
\begin{align*}
 M^2 = 1\;\text{: }\frac{\beta}{2}r_* - \gpjp\frac{1}{1+ 4\gmjp \frac{\tir}{r}}\;\tir = 0\,,
\end{align*}
or
\begin{subequations}\label{eqn:cpbeta}
\begin{align}
 M^2 = 1\;\text{: }r_* = \frac{1}{\beta} \left[ (2\beta + 1) - (2\beta - 1)\gamma\right]\tir\,.\label{eqn:cpbeta1}
\end{align}
Rewriting (\ref{eqn:cpbeta1}) as,
\begin{align}
 r_* &= \left[ 1 - \frac{1}{\beta}\left(\frac{\beta}{2}-1\right) (\gpj) - \frac32 (\gmj)\right]\,\tir\,,\label{eqn:cpbeta2}
\end{align}
and comparing with (\ref{eqn:cp}), super-radial flow conditions ($\beta>2$) are seen to cause the Parker sonic critical point to move further closer to the sun (see Figure~\ref{fig:r_star_beta}) in a polytropic gas ($\gamma>1$). Furthermore, (\ref{eqn:cpbeta1}) implies that
\begin{align}
 \gamma < \left(\frac{2\beta+1}{2\beta - 1}\right) = \frac53 - \frac43 \left(\frac{\beta-2}{2\beta-1}\right)<\frac53\,,\quad\beta>2\,.\label{eqn:cpbeta3}
\end{align}
\end{subequations}
So, in a super-radial flow, the gas is heated up even more to keep $\gamma$ further below $5/3$.

\subsubsection{Exact Solution}
Upon imposing the boundary condition (\ref{eqn:cpbeta1}), equation (\ref{eqn:mom5beta}) yields the \emph{exact} solution,
\begin{align}
\begin{split}
&-\log{M^2} + \gpmj\log{\left(\frac{1+\gmjp M^2}{\gpjp}\right)}\\
 &= -\left( \frac{(2\beta + 1) - (2\beta - 1)\gamma}{\gmj}\right)\log{\left( \frac{\beta\left(\frac{r}{\tir}\right)}{(2\beta+1) - (2\beta - 1)\gamma}\right)}\\
 & + \gpmj \log{\left( \frac{r + 4\,\gmjp\,\tir}{\frac{2}{\beta}\gpjp\,\tir}\right)}\,.
 \end{split} \label{eqn:exsolbeta}
\end{align}

In the radial-flow limit, $\beta=2$, (\ref{eqn:exsolbeta}) reduces to the polytropic gas solution (\ref{eqn:solex}), and in the isothermal gas limit, $\gamma=1$, (\ref{eqn:exsolbeta}) reduces to (\ref{eqn:exsolisobeta}).

\begin{figure}[htb]
 \centering
 \includegraphics[width=\textwidth]{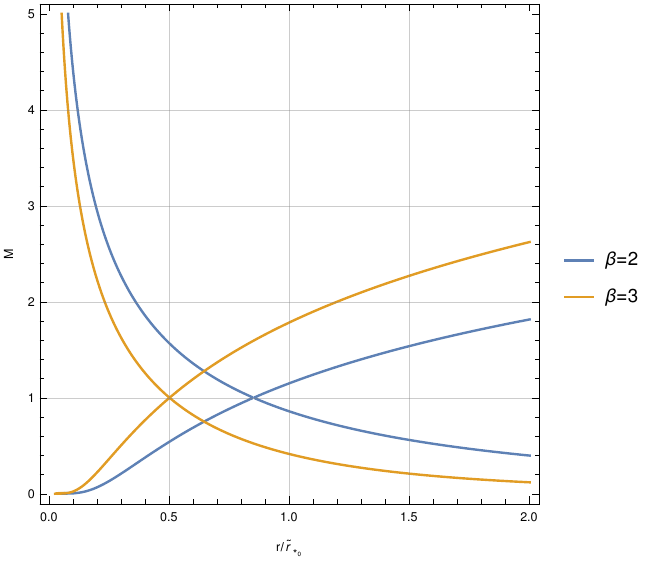}
 \caption{Exact solution (\ref{eqn:exsolbeta}) for a spherical radial expansion ($\beta=2$) and for super-radial expansion ($\beta=3$) of the polytropic solar wind ($\gamma = 1.1$).}
 \label{fig:polytropic_holes}
\end{figure}

The exact solution (\ref{eqn:exsolbeta}) is plotted in Figure~\ref{fig:polytropic_holes}, for different values of the super-radiality parameter $\beta$. Observe the enhanced acceleration of the coronal-hole outflows, which is plausible because of the enhanced flow expansion for such flows.

\subsubsection{Near-sun Regime}
For $r/\tir\ll1$, (\ref{eqn:exsolbeta}) reduces to,
\begin{subequations}\label{eqn:exsolbeta_ns}
\begin{align}
 \log{M^2} &\approx \left[ \frac{(2\beta + 1) - (2\beta - 1)\gamma}{\gmj}\right] \log{r}\,,\label{eqn:exsolbeta_ns1}
\end{align}
from which,
\begin{align}
 M^2 &\sim r^{\frac{(2\beta + 1)-(2\beta - 1)\gamma}{\gmj}}\sim r^{\frac{\pmtg}{\gmj}-2(\beta-2)}\,.\label{eqn:exsolbeta_ns2}
\end{align}
\end{subequations}
On comparing with (\ref{eqn:nearsun2}), super-radial flow conditions, $\beta>2$, are seen to cause an enhancement of flow acceleration near the sun.

\subsubsection{Far-sun Regime}
For $r/\tir\gg1$, (\ref{eqn:exsolbeta}) reduces to,
\begin{subequations}\label{eqn:exsolbeta_fs}
\begin{align}
 \left(-1 + \gpmj\right)\log{M^2} &\approx \left( \frac{(2\beta - 1)\gamma - (2\beta+1)}{\gmj} + \gpmj\right)\log{r}\,,\label{eqn:exsolbeta_fs1}
\end{align}
from which,
\begin{align}
 M^2 &\sim r^{\beta(\gmj)}\,.\label{eqn:exsolbeta_fs2}
\end{align}
\end{subequations}
On comparing with (\ref{eqn:farsun2}), super-radial flow conditions, $\beta>2$, are seen to cause an enhancement of flow acceleration also far away from the sun.

\subsubsection{Transonic Regime}
Near the Parker sonic critical point, $r=r_*=\frac{(2\beta+1)-(2\beta-1)\gamma}{\beta}\;\tir$, we write,
\begin{align}
 r &\equiv \frac{1}{\beta}\left[ (2\beta + 1)-(2\beta-1)\gamma\right]\,\tir + x\,,\quad M^2 \equiv 1+ y\,.
\end{align}
(\ref{eqn:exsolbeta}) then reduces to,
\begin{align*}
 &-\log{(1+y)} + \gpmj\log{\left( 1 + \gmpj\,y \right)}\nonumber \\
 &= -\left( \frac{(2\beta + 1) - (2\beta -1)\gamma}{\gmj}\right)\log{\left[ 1 + \left(\frac{\beta}{(2\beta+1)-(2\beta-1)\gamma}\right)\,\frac{x}{\tir} \right]}\nonumber\\
 &+ \gpmj \log{\left(1 + \frac{\beta}{\gpj}\frac{x}{\tir}\right)}\,,
\end{align*}
and Taylor expanding in powers of $y$ and $x/\tir$, we obtain,
\begin{subequations}\label{eqn:exsolbeta_ts}
\begin{align}
 y &\approx \left( \frac{\beta^3}{(2\beta + 1) - (2\beta-1)\gamma}\right)^{1/2}\,\frac{x}{\tir}\,,\label{eqn:exsolbeta_ts1}
\end{align}
which may be rewritten as,
\begin{align}
 y &\approx \left( \frac{\beta^3}{(\pmtg) - 2(\beta-2)(\gamma-1)}\right)^{1/2}\,\frac{x}{\tir}\,,\label{eqn:exsolbeta_ts2}
\end{align}
or
\begin{align}
 M^2 - 1 &\approx \left( \frac{\beta^3}{(\pmtg) - 2(\beta-2)(\gamma-1)}\right)^{1/2} \left(\frac{r}{\tir} - \frac{(2\beta+1) - (2\beta-1)\gamma}{\beta}\right)\,.\label{eqn:exsolbeta_ts3}
\end{align}
\end{subequations}
(\ref{eqn:exsolbeta_ts}) describe the asymptotes near the Parker sonic critical point ($M^2 = 1$, $r=\frac{\btpm}{\beta}\tir$) to the hyperbolas given by (\ref{eqn:exsolbeta}). This result, as in the radial-flow case ($\beta=2$), implies that the Parker sonic critical point is of $X$-type, which facilitates a smooth transition from subsonic to supersonic wind flow through the transonic regime ($M\approx1$). Observe the enhancement in the flow acceleration past the Parker sonic critical point caused by super-radial flow conditions ($\beta>2$).

It may be noted that the above asymptotic results (\ref{eqn:exsolbeta_ns}), (\ref{eqn:exsolbeta_fs}), and (\ref{eqn:exsolbeta_ts}), for $r/\tir \ll 1$, $r/\tir \gg1$ and $r\approx \frac{\btpm}{\beta}\tir$ may be deduced directly form equation (\ref{eqn:mom5beta}) governing the super-radial, polytropic gas flow (see Appendix \ref{sec:app2}).

\section{Discussion}
In recognition of the impairment of the isothermal gas assumption in Parker's solar wind model due to extended active heating in the corona, we have given a detailed and systematic investigation of polytropic gas effects in Parker's model. We have presented a viable equation governing the acceleration of the solar wind of a polytropic gas, and have given its analytical and numerical solutions, and have exhibited their asymptotic analytic  properties (i) near the sun, (ii) far away from the sun, (iii) near the Parker sonic critical point (where the wind speed is equal to the speed of sound in the wind). On the other hand, the bulk of the solar wind is found to emerge from the coronal holes, so we have given a detailed and systematic investigation of coronal-hole polytropic gas outflows. We have modeled coronal-hole outflow by considering a single radial stream tube and invoked phenomenological considerations to represent its rapidly diverging flow geometry. We have given \emph{exact analytical} and numerical solutions for this outflow and exhibited their asymptotic analytic properties in the three flow regimes listed above. We have found that, in general, the polytropic effects cause the Parker sonic critical point to move closer to the sun than that for the case with isothermal gas. Furthermore, the flow acceleration has been found to exhibit (even for an infinitesimal deviation from isothermality of the gas) a power-law behavior rather than an exponential-law behavior near the sun or a logarithmic-law behavior far away from the sun, thus implying a certain robustness of the power-law behavior for the flow acceleration. This also signifies that the isothermal limit, $\gamma\to1$, is a \emph{singular} limit of the polytropic gas, $\gamma>1$, so the polytropic gas results are not obtainable via a Taylor series expansion about the isothermal case. Parker sonic critical point has been shown to continue to be of $X$-type, hence, facilitating a smooth transition from subsonic to supersonic wind flow through the transonic regime. Our analytical and numerical solutions for coronal-hole outflows show that the super-radiality of the stream tube causes the Parker sonic critical point to move further down in the corona, and the gas to become more diabatic (the polytropic exponent $\gamma$ drops further below $5/3$), and the flow acceleration to be enhanced further in agreement with numerical calculations of \textcite{Kopp1976}.

It may be mentioned that \emph{in situ} observations of the solar winds have indicated that the solar wind, contrary to what is assumed in the steady models, is far from being steady and structureless. However, the large-scale behavior of the solar wind, on the average, its local noisiness, notwithstanding, has been found to be close to Parker’s solar wind solution. This endows Parker’s solar wind solution with a certain robustness and indicates an ability to sustain itself against small disturbances acting on this system. Indeed, \textcite{Parker1960} proposed that Parker’s solution exhibits an intrinsic stability like a “\emph{stable attractor}” of this dynamical system \parencite{Cranmer2019}.  The stability of Parker’s solar wind solution with respect to linearized perturbations was investigated with the solar corona in the subcritical region approximated by a static atmosphere \parencite{Parker1966,Shivamoggi2023}. The inclusion of a solar wind flow in the basic state by \textcite{Carovillano1966,Jockers1968} indicated that this problem possesses a singularity at Parker’s sonic critical point, which makes this problem ill-posed. \textcite{Shivamoggi2024} gave a recipe to regularize this singularity by dealing with the corresponding non-linear problem. However, this treatment is restricted to isothermal gas flows,  and hence needs to be extended to cover polytropic gas flows in a future work.

\section{Acknowledgements}
This work was carried out during BKS's sabatical leave at California Institute of Technology. BKS thanks Professor Shrinivas Kulkarni for his enormous hospitality and valuable remarks and Drs. Elias Most and Reem Sari for the helpful discussions. BKS is thankful to Professors Earl Dowell and Katepalli Screenivasan for their helpful remarks and suggestions.

\begin{appendices}
\setcounter{equation}{0}
\renewcommand{\theequation}{\thesection.\arabic{equation}}
 \section{Polytropic Effects}\label{sec:app1}
 \subsection{Near-sun Regime}
 For $r/\tir\ll1$, equation (\ref{eqn:mom5}) reduces to,
 \begin{align}
  -\frac{1}{M^2}\ddd{M^2} &\approx \frac{4}{r^2} \left( r - \frac{\tir}{4} \gpmj \frac{r}{\tir}\right)\,,
 \end{align}
or
\begin{align}
 \ddd{\left( \log{M^2} \right)} &\approx \frac{\pmtg}{\gmj}\,\frac{1}{r}\,,
\end{align}
from which
\begin{align}
 M^2 &\sim r^{\frac{\pmtg}{\gmj}}\,,
\end{align}
in agreement with (\ref{eqn:nearsun2}).

\subsection{Far-Sun Regime}
For $r/\tir\gg1$, equation (\ref{eqn:mom5}) reduces to,
 \begin{align}
  \frac{2}{\gmj}\frac{1}{M^2} \ddd{M^2} &\approx \frac{4}{r}\,,
 \end{align}
from which,
\begin{align}
 M^2 &\sim r^{2(\gmj)}\,,
\end{align}
in agreement with (\ref{eqn:farsun2}).

\subsection{Transonic Regime}
Near the Parker sonic critical point $r = r_* = \pmtgp\,\tir$, we write,
\begin{align}
 r &\equiv \pmtgp\,\tir  + x\,,\quad M^2 \equiv 1 + y\,,
\end{align}
equation (\ref{eqn:mom5}) then reduces to,
\begin{align}
 \begin{split}
 &\frac{y}{(1+y)\left[ 1 + \gmjp (1+y) \right]}\ddd{y} \\
 &= \frac{4}{\left( \pmtgp\,\tir + x\right)^2}\left[ \left( \pmtgp\,\tir + x\right) - \gpjp\frac{\tir}{1+ \frac{2(\gmj)\tir}{\pmtgp\,\tir + x} }\right]\,,
 \end{split}
\end{align}
and Taylor expanding in powers of $y$ and $x/\tir$, we obtain,
\begin{align}
 \begin{split}
 \frac{2y}{\gpj} \ddd{y} &\approx \frac{16}{(\pmtg)^2\,\tir}\,\Biggl[ \left( \pmtgp - \gpjp\frac{1}{1+4 \frac{\gmj}{\pmtg}}\right) \\
 &+\left( \pmtgp - 2 \frac{\gamma^2 - 1}{\pmtg} \frac{1}{\left(1+4\frac{\gmj}{\pmtg}\right)^2}\right)\frac{2}{\pmtg}\frac{x}{\tir}\Biggr]\,,
 \end{split}
\end{align}
or
\begin{align}
 y\ddd{y} &\approx \frac{8}{\pmtg}\;\frac{x}{\tir^2}\,,
\end{align}
from which we obtain
\begin{align}
 y &\approx \sqrt{\frac{8}{\pmtg}}\;\frac{x}{\tir}\,,
\end{align}
in agreement with (\ref{eqn:tran1}).

\section{Coronal Hole Flows}\label{sec:app2}
\setcounter{equation}{0}
\subsection{Near-sun Regime}
For $r/\tir\ll1$, equation (\ref{eqn:mom5beta}) reduces to,
\begin{align}
 -\frac{1}{M^2}\ddd{M^2} &\approx \frac{4}{r^2} \left( \frac{\beta}{2}\,r - \frac{\tir}{4}\gpmj\,\frac{r}{\tir}\right)\,,
\end{align}
or
\begin{align}
 \ddd{\left( \log{M^2}\right)} &\approx \frac{\btpm}{\gmj}\,\frac{1}{r}\,,
\end{align}
from which,
\begin{align}
 M^2 &\sim r^{\frac{\btpm}{\gmj}}\,,
\end{align}
in agreement with (\ref{eqn:exsolbeta_ns2}).

\subsection{Far-sun Regime}
For $r/\tir\gg1$, equation (\ref{eqn:mom5beta}) reduces to,
\begin{align}
 \frac{2}{\gmj}\frac{1}{M^2}\ddd{M^2} &\approx \frac{2\beta}{r}\,,
\end{align}
from which,
\begin{align}
 M^2 &\sim r^{\beta(\gmj)}\,,
\end{align}
in agreement with (\ref{eqn:exsolbeta_fs2}).

\subsection{Transonic Regime}
Near the Parker sonic critical point $r=r_* = \left(\frac{\btpm}{\beta}\right)\,\tir$, we write,
\begin{align}
 r &\equiv \frac{\btpm}{\beta}\,\tir + x\,,\quad M^2 \equiv 1 + y\,,
\end{align}
equation (\ref{eqn:mom5beta}) then reduces to,
\begin{align}
\begin{split}
 &\frac{y}{(1+y)\left[ 1 + \gmjp (1+y)\right]}\ddd{y} \\
 &= \frac{4}{\left( \left(\frac{\btpm}{\beta}\right)\,\tir + x \right)^2}\,\Biggl[ \frac{\beta}{2} \left( \left(\frac{\btpm}{\beta}\right)\,\tir + x \right) \\
 &- \gpjp\frac{\tir}{1 + \frac{2(\gmj)\tir}{\left(\frac{\btpm}{\beta}\right)\,\tir + x} } \Biggr]\,,
\end{split}
\end{align}
and Taylor expanding in powers of $y$ and $x/\tir$, we obtain,
\begin{align}
    \begin{split}
    \frac{2y}{\gpj} \ddd{y} &\approx \frac{4}{\left(\frac{\btpm}{\beta}\right)^2\,\tir^2}\times\\
    &\times\left[ \frac{\beta}{2} - \frac{\gamma^2 - 1}{\left\{\frac{1}{\beta}\left[ \btpm \right]+2(\gmj) \right\}^2}\right]\,x\,,
    \end{split}
\end{align}
or
\begin{align}
  y\,\ddd{y} &\approx \left[ \frac{\beta^3}{(2\beta+1) - (2\beta-1)\gamma}\right]\,\frac{x}{\tir^2}\,,
\end{align}
from which we obtain
\begin{align}
 y &\approx \left[ \frac{\beta^3}{\btpm} \right]^{1/2} \;\frac{x}{\tir}\,,
\end{align}
in agreement with (\ref{eqn:exsolbeta_ts1}).

\end{appendices}
\printbibliography

\end{document}